\documentclass[prd,preprint,superscriptaddress,showpacs,nofootinbib,%
tightenlines]{revtex4}
\usepackage{epsfig}
\newcommand{\slashed}{\not\hspace{-0.7mm}}
\begin{document}
\title{Consistency of the $\pi\Delta$ interaction in chiral perturbation
theory}
\author{N.~Wies}
\affiliation{Institut f\"ur Kernphysik, Johannes
Gutenberg-Universit\"at, D-55099 Mainz, Germany}
\author{J.~Gegelia}
\affiliation{Institut f\"ur Kernphysik, Johannes
Gutenberg-Universit\"at,  D-55099 Mainz,
Germany}
\affiliation{
High Energy Physics Institute, Tbilisi State University,
Tbilisi, Georgia}
\author{S.~Scherer}
\affiliation{Institut f\"ur Kernphysik, Johannes
Gutenberg-Universit\"at, D-55099 Mainz, Germany}
\date{\today}

\begin{abstract}
   We analyze the constraint structure of a spin-3/2 particle
interacting with a pseudoscalar.
   Requiring the self consistency of the considered effective
field theory imposes restrictions on the possible interaction
terms.
   In the present case we derive two constraints among the
three lowest-order $\pi\Delta$ interaction terms.
   From these constraints we find that the total Lagrangian
is invariant under the so-called point transformation.
   On the other hand, demanding the invariance under the point
transformation alone is less stringent and produces only classes
of relations among the coupling constants.
\end{abstract}
\pacs{
12.39.Fe,
14.20.Gk
}

\maketitle
\section{Introduction}

   The $\Delta(1232)$ resonance plays an important role in the phenomenological
description of low- and medium-energy processes such as pion-nucleon
scattering, electromagnetic pion production, Compton scattering etc..
   This is due to the rather small mass gap between the $\Delta(1232)$ and
the nucleon, the strong coupling of the $\Delta(1232)$ to the $\pi N$ channel,
and its relatively large photon decay amplitudes.
   In an effective-field-theory (EFT) approach \cite{Weinberg:1979kz}
to, say, the single-nucleon sector one encounters two possibilities.
   In conventional baryon chiral perturbation theory (BChPT)
\cite{Gasser:1988rb,Jenkins:1991jv,Bernard:1992qa} (see, e.g.,
Refs \cite{Scherer:2002tk,Scherer:2005ri} for an introduction) one
is restricted to the threshold regime (of pion production) and the
dynamical effects due to the $\Delta(1232)$ are encoded in the
values of the low-energy constants of the most general effective
Lagrangian.
   Alternatively, one can try to include the $\Delta(1232)$ as an
explicit dynamical degree of freedom.
   In doing so, one hopes to improve the convergence behavior of the
chiral expansion by reordering important terms which in an
ordinary chiral expansion would show up at higher orders.
   Moreover, if one succeeds in defining a suitable power-counting scheme
one may even be able to perform calculations of processes which
involve center-of-mass energies covering the resonance region.
   Clearly, there is a strong motivation for taking the $\Delta(1232)$ as an
explicit dynamical degree of freedom into account and this issue
has already attracted considerable attention for quite some time.
   So far, most of the calculation have been performed in the heavy-baryon
framework
\cite{Jenkins:1991es,Butler:1992pn,Butler:1993ht,Butler:1993ar,%
Butler:1993ej,Banerjee:1995wz,Hemmert:1997ye,Gellas:1998wx,%
Kao:1998xk,Puglia:2000jy,Gail:2005gz}.
   On the other hand, more recently several methods have been devised
to obtain renormalization schemes leading to a simple and consistent
power counting in a manifestly Lorentz-invariant approach
of BChPT
\cite{Tang:1996ca,Ellis:1997kc,Becher:1999he,Lutz:1999yr,%
Gegelia:1999gf,Gegelia:1999qt,Lutz:2001yb,Fuchs:2003qc}.
   It is thus natural to investigate a consistent Lorentz-invariant
formulation including spin-3/2 degrees of freedom (see, e.g.,
Refs.~\cite{Pascalutsa:2002pi,Bernard:2003xf,
Pascalutsa:2004je,Walker-Loud:2005bt,
Bernard:2005fy,Hacker:2005fh,Pascalutsa:2005es,Pascalutsa:2005nd,
Semke:2005sn,Pascalutsa:2005vq}).

   BChPT with explicit $\Delta$ degrees of freedom ($\Delta$ChPT)
is a field theory of a system with constraints.
   Therefore, one encounters the highly non-trivial problem
of a {\it consistent interaction of higher-spin fields} (see,
e.g.,
Refs.~\cite{Dirac:1936tg,Fierz:1939ix,Johnson:1960vt,Velo:1969bt}).
   In a Lorentz-invariant formulation of a field
theory involving particles of higher spin ($s\geq 1$), one
necessarily introduces unphysical degrees of freedom
\cite{Rarita:1941mf,case:1955}.
   Therefore one has to impose constraints which specify the physical
degrees of freedom.
   To write down interaction terms which lead to the correct number
of physical degrees of freedom has proven to be a difficult
problem.
   There are various suggestions for constructing consistent
interactions involving spin-3/2 particles (see, e.g.,
Refs.~\cite{Nath:1971wp,Tang:1996sq,Pascalutsa:1998pw,
Pascalutsa:1999zz,Deser:2000dz,Pascalutsa:2000kd,
Pilling:2004wk,Pilling:2004cu,Jahnke:2006nj}).
   In this context we note that the problems
showing up only for large field configurations are not relevant to
low-energy EFTs because these deal with small fluctuations of
field variables around the vacuum.
   For larger field configurations the higher-order terms (infinite in number)
generate contributions to physical quantities which are no longer
suppressed by powers of small expansion parameters.
   Therefore, for large fluctuations the conclusions drawn from an analysis
of a finite number of terms of the effective Lagrangian cannot be
trusted.
   On the other hand, for small fluctuations around the vacuum one requires
that the theory describes the right number of degrees of freedom
in a self-consistent way.
   The interaction terms can be analyzed order by order in a small parameter
expansion.
   Such an analysis leads to non-trivial constraints on the possible
interactions.

   In the present paper we consider the leading-order interaction terms of
the pion with the $\Delta(1232)$ in low-energy effective field
theory---$\Delta$ChPT.
   We derive consistency conditions for the
$\pi\Delta$-interaction terms by analyzing the structure of the
constraints using the canonical (Hamilton) formalism.
   For reasons of simplicity we suppress the isospin degree of freedom.

\section{Properties of the Lagrangian for a spin-3/2 system}
\subsection{The  free Lagrangian of a spin-3/2 system}
\label{freiesDelta}
   Fields with spin 3/2 can be described via the
Rarita-Schwinger formalism, where the field is represented by a
vector spinor $\psi^{\mu}$ \cite{Rarita:1941mf}. The most general
free Lagrangian reads \cite{case:1955}
\begin{equation}
\mathcal{L}_{\frac{3}{2}}=
\bar{\psi}^{\alpha}\,\Lambda^A_{\alpha\beta}\,\psi^{\beta}\,,
\label{LfreiA}
\end{equation}
where
\begin{equation}
\Lambda^A_{\alpha\beta}=-[(i\slashed{\partial}-m)\,g_{\alpha\beta}+i
A\,(\gamma_{\alpha}\partial_{\beta}+\gamma_{\beta}\partial_{\alpha})
+\frac{i}{2}(3A^2+2A+1)\gamma_{\alpha}\slashed{\partial}\gamma_{\beta}
+m(3A^2+3A+1)\,\gamma_{\alpha}\gamma_{\beta}]\,,\label{lambdaA}
\end{equation}
with $A\neq -1/2$ an arbitrary real parameter.

   The generalization for an arbitrary space-time dimension $n$ is
(see, e.g., Ref.~\cite{Pilling:2004cu})
\begin{equation}
\mathcal{L}_{\frac{3}{2}}=
\bar{\psi}^{\alpha}\,\Lambda^{A,\,n}_{\alpha\beta}\,\psi^{\beta}\,,
\label{LfreiAndim}
\end{equation}
where
\begin{eqnarray}
\Lambda^{A,\,n}_{\alpha\beta}&=&-\left\{(i\slashed{\partial}-m)\,
g_{\alpha\beta}+i
A\,(\gamma_{\alpha}\partial_{\beta}+\gamma_{\beta}\partial_{\alpha})
+\frac{i}{n-2}\left[(n-1)A^2+2A+1\right]
\gamma_{\alpha}\slashed{\partial}\gamma_{\beta}\right.
\nonumber\\
&&\left.+\frac{m}{(n-2)^2}\left[
n(n-1)A^2+4(n-1)A+n\right]\,\gamma_{\alpha}\gamma_{\beta}\right\}\,,
\qquad\qquad n\neq2\,.\label{lambdafreiAn}
\end{eqnarray}
   In the special case of $A=-1$, Eq.\ (\ref{LfreiAndim}) does not
explicitly depend on $n$.

\subsection{Point invariance}

   The free Lagrangian of Eq.\ (\ref{LfreiAndim}) is invariant under
the set of transformations
\begin{eqnarray}
\psi_{\mu}&\rightarrow&\psi_{\mu}
+\frac{4a}{n}\gamma_{\mu}\gamma_{\nu}\psi^{\nu}\label{punkt1}\,,\\
A&\rightarrow&\frac{An-8a}{n(1+4a)}\,, \qquad\qquad a\neq
-\frac{1}{4}\,, \label{punkt2}
\end{eqnarray}
which are often referred to as a point transformation.
   The invariance under the point transformation guarantees that the
physical quantities do not depend on $A$
\cite{Nath:1971wp,Tang:1996sq}, provided that the interaction
terms are also invariant under the point transformation.
   We will not {\em impose} the last condition but will rather
{\em obtain} it as a consequence of consistency in the sense of
having the right number of degrees of freedom.

\section{General considerations}
\label{allgbetr}
   In this section we outline the method of analyzing systems with
constraints of the second class starting with a finite number of degrees
of freedom.
   For a more detailed description see, e.g., Refs.\
\cite{Dirac,Gitman:1990qh,teitelboim}.
   Let us consider a classical system with $N$
degrees of freedom $q_i$ and velocities $\dot q_i=dq_i/dt$
described by the Lagrangian $L(q,\dot q)$.
   Here, we assume that $L$ contains the  $\dot q$'s at the most quadratically.
   The Hamiltonian is obtained using the Legendre transform
\begin{equation}
H(q,p)=\sum_{i=1}^N p_i\dot q_i-L(q,\dot q)\,,\label{legendre}
\end{equation}
where the $p_i$ are the canonical momenta defined by
\begin{equation}
p_i\equiv\frac{\partial L(q,\dot q)}{\partial \dot q_i}\,,\quad
i=1,\ldots, N\,.\label{kanonimp}
\end{equation}
   Since the Hamiltonian is a function of $q$ and $p$, the velocities
$\dot q_i$ have to be replaced using Eq.\ (\ref{kanonimp}).
   If, according to Eq.~(\ref{kanonimp}), this is not possible because
\begin{equation}
\det \mathrm A=0\,,\quad\textnormal{with}\quad\mathrm
A_{ij}=\frac{\partial p_i}{\partial \dot q_j}\,,
\end{equation}
we are dealing with a singular system.
   With a suitable change of coordinates, the Lagrangian can be written as a
linear function of the unsolvable new velocities $\dot q'_i$.
   In the following the new coordinates are again denoted by $q_i$.
   Let the unsolvable $\dot q_i$ be the first $n$ velocities
$\dot q_1,\ldots, \dot q_n$.
   The so-called primary constraints occur as follows.
   The Lagrangian $L$ can be written as
\begin{equation}
L(q,\dot q)=\sum_{i=1}^n F_i(q) \dot q_i+G(q,\dot
q_{n+1},\ldots,\dot q_N)\,
\end{equation}
from which we obtain as the canonical momenta
\begin{equation}
p_i=\left\{\begin{array}{ll}F_i(q)&\textnormal{for } i=1,\ldots
,n\,,\\\frac{\partial G(q,\dot q_{n+1},\ldots,\dot q_N)}{\partial
\dot q_i}&\textnormal{for
}i=n+1,\ldots,N\,.\end{array}\right.\label{kanonimp'}
\end{equation}
   The first part of Eq.\ (\ref{kanonimp'}) can be reexpressed
in terms of the relations
\begin{equation}
\theta_i(q,p)=p_i-F_i(q)=0\,,\qquad i=1,\ldots,n\,,
\end{equation}
which are referred to as the primary constraints.
   Using Eq.\ (\ref{legendre}), we consider the Hamiltonian
\begin{equation}
H(q,p)=\sum_{j=n+1}^N p_j\dot q_j(p,q)
-G(q,\dot q_{n+1}(p,q),\ldots,\dot q_N(p,q))
+\sum_{i=1}^n\lambda_i \theta_i(q,p)\,,
\end{equation}
where the $\lambda$'s are Lagrange multipliers taking care of the
primary constraints and the $\dot q_i(p,q)$ are the solutions to
Eq.~(\ref{kanonimp'}) for $i=n+1,\ldots,N$.
   We determine the $\lambda$'s by using the condition that the
constraints have to be conserved in time.
   The time evolution of the primary constraints $\theta_i$ is given
by the Poisson bracket with the Hamiltonian
so that the condition for conservation in time reads
\begin{equation}
\{\theta_i,H\}\stackrel{!}{=}0\,.
\end{equation}
   Either all the $\lambda$'s can be determined from these equations, or
new constraints arise.
   The number of these secondary constraints corresponds to the number
of the $\lambda$'s (or linear combinations thereof) which could not
be determined.
   Again we demand the conservation in time of these (new) constraints and
try to solve the remaining $\lambda$'s from these equations, etc..
   The number of physical degrees of freedom is given by the initial number of
degrees of freedom minus the number of constraints.
   In order for a theory to be consistent, the chain of new
constraints has to terminate such that at the end of the procedure
the correct number of degrees of freedom has been generated.

\section{Consistency conditions for the $\pi\Delta$ interaction terms}
\label{quant}
   In this section we will determine the Hamiltonian for pions and
deltas in analogy to the discussion above.
   The consistency of the interaction terms in the Lagrangian is only
guaranteed, if the correct number of degrees of freedom is generated.
   Taking the fields $\psi^{\mu}$ and $\psi^{\mu\dag}$ and the
canonical momenta $\pi^\mu_\psi$ and $\pi^\mu_{\psi^\dagger}$ as the
independent variables we have in total
$2\times2\times4\times4=64$ components, so 48 constraints are needed to
obtain the right number of degrees of freedom for a
spin-3/2 system.
   This will generate conditions among the coupling constants $g_1$, $g_2$,
and $g_3$ of the original Lagrangian, as will be shown in the
following.

\subsection{Constraint analysis in four dimensions}
   The total leading-order Lagrangian $\mathcal
L$ is given by
\begin{equation}
\mathcal L=\mathcal L_0+\mathcal{L}_{\frac{3}{2}}+\mathcal
L_{\pi\Delta}\,,\label{Lges}
\end{equation}
where $\mathcal{L}_{0}$ denotes the free-pion
Lagrangian
\begin{equation}
\mathcal{L}_0
=\frac{1}{2}\partial_{\mu}\phi\partial^{\mu}\phi-\frac{1}{2}M^2\phi^2\,,
\end{equation}
$\mathcal{L}_{\frac{3}{2}}$ is given in Eq.~(\ref{LfreiA}), and
the leading-order $\pi\Delta$-interaction terms read
\cite{Hemmert:1997ye}\footnote{
Reference \cite{Hemmert:1997ye} discusses the most general {\em chiral}
spin-3/2 Lagrangian which, in particular, implies that one deals
with pion and  $\Delta$  isospin triplets and quadruplets,
respectively.
   We have simplified the discussion by neglecting the isospin degree
of freedom, because the present results do not depend on the isospin
structure. Moreover, the present coupling constants $g_i$ correspond
to $-g_i/F$ in the full chiral case, where $F$ denotes the pion-decay
constant in the chiral limit.}
\begin{equation}
\mathcal{L}_{\pi\Delta}=
-\bar{\psi}^{\mu}\biggl[\frac{g_1}{2}g_{\mu\nu}
\gamma^{\alpha}\gamma_5\partial_{\alpha}\phi
+\frac{g_2}{2}(\gamma_{\mu}\partial_{\nu}\phi
+\partial_{\mu}\phi\gamma_{\nu})\gamma_5
+\frac{g_3}{2}\gamma_{\mu}
\gamma^{\alpha}\gamma_5\gamma_{\nu}\partial_{\alpha}\phi\biggr]\psi^{\nu}\,.
\label{piDWW}
\end{equation}
   The values of the coupling constants $g_i$ are {\it not}
restricted by symmetries imposed on the most general effective
Lagrangian.
   For convenience we will use a manifestly hermitian expression for
$\mathcal{L}_{\frac{3}{2}}$ differing from Eq.~(\ref{LfreiA}) by an
irrelevant total derivative:
\begin{eqnarray}
\mathcal{L}_{\frac{3}{2}}&=&-\bar{\psi}^{\alpha}
\biggl[\left(\frac{i}{2}(\overrightarrow{\slashed{\partial}}
-\overleftarrow{\slashed{\partial}})-m\right)g_{\alpha\beta}
+\frac{iA}{2}\left(\gamma_{\alpha}\overrightarrow{\partial}_{\beta}
-\gamma_{\alpha}\overleftarrow{\partial}_{\beta}
+\gamma_{\beta}\overrightarrow{\partial}_{\alpha}
-\gamma_{\beta}\overleftarrow{\partial}_{\alpha}\right)\nonumber\\
&&+\frac{i}{4}(3A^2+2A+1)
\left(\gamma_{\alpha}\overrightarrow{\slashed{\partial}}\gamma_{\beta}
-\gamma_{\alpha}\overleftarrow{\slashed{\partial}}\gamma_{\beta}\right)
+m(3A^2+3A+1)\gamma_{\alpha}\gamma_{\beta}\biggr]\psi^{\beta}\,.
\end{eqnarray}

   We will express the Hamiltonian corresponding to Eq.~(\ref{Lges})
as the sum of four terms,
\begin{equation}
\mathcal{H}=\mathcal{H}_1+\mathcal{H}_{2}+\mathcal{H}_{3}+\mathcal{H}_{4}\,.
\end{equation}
   To that end we first calculate the canonical momenta.
   The canonical momentum field $\pi$ corresponding to the pion field is
given by
\begin{equation}
\pi=\frac{\partial\mathcal
L}{\partial\dot{\phi}}
=\frac{\partial\mathcal
L_0}{\partial\dot{\phi}}+\frac{\partial\mathcal
L_{\pi\Delta}}{\partial\dot{\phi}}
=\dot{\phi}+\mathcal F\,,
\end{equation}
where
\begin{eqnarray}
\mathcal F&=&\left(-\frac{g_1}{2}-g_2+\frac{g_3}{2}\right)
\bar{\psi}_0\gamma_0\gamma_5\psi_0
+\left(\frac{g_2}{2}-\frac{g_3}{2}\right)
\bar{\psi}_i\gamma_i\gamma_5\psi_0\nonumber\\
&&+\frac{g_1}{2}\bar{\psi}_i\gamma_0\gamma_5\psi_i
+\left(\frac{g_2}{2}-\frac{g_3}{2}\right)
\bar{\psi}_0\gamma_i\gamma_5\psi_i
+\frac{g_3}{2}\bar{\psi}_i\gamma_0\gamma_i\gamma_5\gamma_j\psi_j\,.
\end{eqnarray}
   We define $\mathcal{H}_1$ as
\begin{equation}
\mathcal H_1=\pi\dot{\phi}-\mathcal L_0
=\frac{1}{2}(\pi^2-\mathcal F^2)
+\frac{1}{2}(\partial_i\phi)^2+\frac{1}{2}M^2\phi^2\,.
\end{equation}
   The second term, $\mathcal{H}_2$, is defined by
\begin{eqnarray}
\mathcal H_{2}&=&-\mathcal L_{\pi\Delta}\nonumber\\
&=&-\mathcal F(\pi-\mathcal F)
+\frac{g_1}{2}\left(\bar{\psi}_j\gamma_i\gamma_5\partial_i\phi\psi_j
-\bar{\psi}_0\gamma_i\gamma_5\partial_i\phi\psi_0\right)\nonumber\\
&&+\frac{g_2}{2}\left(-\bar{\psi}_0\gamma_0\partial_i\phi\gamma_5\psi_i
+\bar{\psi}_j\gamma_j\partial_i\phi\gamma_5\psi_i
-\bar{\psi}_i\gamma_0\partial_i\phi\gamma_5\psi_0
+\bar{\psi}_i\gamma_j\partial_i\phi\gamma_5\psi_j\right)\nonumber\\
&&+\frac{g_3}{2}\left(-\bar{\psi}_0\gamma_i\gamma_5\partial_i\phi\psi_0
+\bar{\psi}_0\gamma_0\gamma_i\gamma_5\gamma_j\partial_i\phi\psi_j
+\bar{\psi}_j\gamma_j\gamma_i\gamma_5\gamma_0\partial_i\phi\psi_0
-\bar{\psi}_j\gamma_j\gamma_i\gamma_5\gamma_k\partial_i\phi\psi_k\right)\,.
\nonumber\\
\end{eqnarray}
   Because of the fermionic nature of the spin-3/2 field,
$\psi^{\mu}$ and $\psi^{\mu\dag}$ are Grassmann fields.
   We define the corresponding momenta as
\begin{eqnarray}
\pi^{\mu}_{\psi}&=&\frac{\partial^R\mathcal
L}{\partial(\partial_0\psi_{\mu})}\nonumber\,,\\
\pi^{\mu}_{\psi^{\dag}}&=&\frac{\partial^L\mathcal
L}{\partial(\partial_0{\psi^{\dag}}_{\mu})}\,,
\end{eqnarray}
where $\partial^R$ and $\partial^L$ denote the right and the left
derivatives.
   We define
\begin{equation}
\mathcal
H_{3}=\pi^\mu_{\psi}\dot{\psi}_{\mu}
+\dot{\psi}_{\mu}^{\dag}\pi^{\mu}_{\psi^{\dag}}
-\mathcal L_{\frac{3}{2}}\,,
\end{equation}
where the time and space components of $\pi^\mu_\psi$ are given by
\begin{eqnarray}
\pi^0_{\psi}&=&\frac{i}{4}(3A^2+4A+1)\bar{\psi}_i\gamma_i
-\frac{3}{4}i(A^2+2A+1)\bar{\psi}_0\gamma_0\,,\\
\pi^i_{\psi}&=&\frac{i}{2}\bar{\psi}_i\gamma_0
+\frac{i}{4}(3A^2+4A+1)\bar{\psi}_0\gamma_i
-\frac{i}{4}(3A^2+2A+1)\bar{\psi}_j\gamma_j\gamma_0\gamma_i\,.
\end{eqnarray}
   Analogously,
\begin{eqnarray}
\pi^0_{\psi^{\dag}}&=&-\frac{i}{4}(3A^2+4A+1)\gamma_0\gamma_i\psi_i
+\frac{3}{4}i(A^2+2A+1)\psi_0\,,\\
\pi^i_{\psi^{\dag}}&=&-\frac{i}{2}\psi_i
-\frac{i}{4}(3A^2+4A+1)\gamma_0\gamma_i\psi_0
-\frac{i}{4}(3A^2+2A+1)\gamma_i\gamma_j\psi_j\,.
\end{eqnarray}
   Since $\dot{\psi}_{\mu}$ and $\dot{\psi}^{\dag}_{\mu}$ do not
appear in these equations, constraints are generated, as described
in the previous section.
    These primary constraints are
\begin{eqnarray}
\theta^0_{\psi}&=&\pi^0_{\psi}-\frac{i}{4}(3
A^2+4A+1)\bar{\psi}_i\gamma_i
+\frac{3}{4}i(A^2+2A+1)\bar{\psi}_0\gamma_0\stackrel{!}{=}0\,,\nonumber\\
\theta^i_{\psi}&=&\pi^i_{\psi}-\frac{i}{2}\bar{\psi}_i\gamma_0
-\frac{i}{4}(3A^2+4A+1)\bar{\psi}_0\gamma_i
+\frac{i}{4}(3A^2+2A+1)\bar{\psi}_j\gamma_j\gamma_0\gamma_i
\stackrel{!}{=}0\,,\label{zwangsbed1}
\end{eqnarray}
and analogous constraints $\theta^0_{\psi^\dag}$ and
$\theta^i_{\psi^\dag}$.
   Finally, we define the last piece in terms of the Lagrange multipliers
and constraints as
\begin{equation}
\label{h4} \mathcal H_4= \theta^\mu_\psi\lambda_\mu
+\lambda_\mu^\dagger\theta^\mu_{\psi^\dagger}\,.
\end{equation}
   From the condition that $\theta^0_{\psi}$ etc.~have to be zero
throughout all time we obtain a set of linear equations for the
eight Lagrange multipliers $\lambda$ and $\lambda^\dag$,
where each component $\lambda_\mu$ and $\lambda_\mu^\dagger$
is a four-component object.

   To this end we define the Poisson brackets
\begin{eqnarray}
\{\psi_{0}(\vec x),\pi^{0}_{\psi}(\vec y)\}
&=&\{\psi^{\dag}_{0}(\vec x),\pi^{0}_{\psi^{\dag}}(\vec y)\}
=\delta^3(\vec x-\vec y)\,,\nonumber\\
\{\psi_{i}(\vec x),\pi^{j}_{\psi}(\vec y)\}
&=&\{\psi^{\dag}_{i}(\vec x),\pi^{j}_{\psi^{\dag}}(\vec y)\}
=\delta_{ij}\delta^3(\vec x-\vec y)\,,\nonumber\\
\{\phi(\vec x),P(\vec y)\}&=&\delta^3(\vec x-\vec y)\,,
\end{eqnarray}
where we have omitted the Dirac-spinor indices.
   In addition, we have
\begin{equation}\{A,B\}
=-(-)^{\mathcal P(A)\mathcal P(B)}\{B,A\}\,,
\end{equation}
where $\mathcal P(X)$ takes the value 1 for Grassmann fields and 0 otherwise.
   The remaining Poisson brackets vanish.
   Since the time evolution is governed by the Poisson brackets, we
obtain from demanding the time independence of the primary constraints
\begin{eqnarray}
0&\stackrel{!}{=}&\{\theta^0_{\psi},H\}\nonumber\\
&=&(\mathcal F-\pi) \left[\left(-\frac{g_1}{2}-g_2+\frac{g_3}{2}\right)
\bar{\psi}_0\gamma_0\gamma_5
+\left(\frac{g_2}{2}-\frac{g_3}{2}\right)\bar{\psi}_i\gamma_i\gamma_5\right]
\nonumber\\
&&-\left(\frac{g_1}{2}+\frac{g_3}{2}\right)
\bar{\psi}_0\gamma_i\gamma_5\partial_i\phi
-\frac{g_2}{2}\bar{\psi}_i\gamma_0\gamma_5\partial_i\phi
+\frac{g_3}{2}\bar{\psi}_j\gamma_j\gamma_i\gamma_5\gamma_0\partial_i\phi
\nonumber\\
&&+iA\partial_i\bar{\psi}_i\gamma_0
-\frac{i}{2}(3A^2+2A-1)\partial_i\bar{\psi}_0\gamma_i
-\frac{i}{2}(3A^2+2A+1)\partial_i\bar{\psi}_j\gamma_j\gamma_i\nonumber\\
&&+m\left[3A(A+1)\bar{\psi}_0
-(3A^2+3A+1)\bar{\psi}\gamma_i\gamma_0\right]\nonumber\\
&&+\frac{i}{2}(3A^2+4A+1)\lambda_i^{\dag}\gamma_0\gamma_i
-\frac{3}{2}i(A^2+2A+1)\lambda_0^{\dag},
\label{poisson0}\\
0&\stackrel{!}{=}&\{\theta^i_{\psi},H\}\nonumber\\
&=&(\mathcal F-\pi)\left(\frac{g_1}{2}\bar{\psi}_i\gamma_0\gamma_5
+\frac{g_2-g_3}{2}\bar{\psi}_0\gamma_i\gamma_5
+\frac{g_3}{2}\bar{\psi}_j\gamma_0\gamma_j\gamma_5\gamma_i\right)\nonumber\\
&&+\frac{g_1}{2}\bar{\psi}_i\gamma_j\gamma_5\partial_j\phi
+\frac{g_2}{2}\left(-\bar{\psi}_0\gamma_0\gamma_5\partial_i\phi
+\bar{\psi}_j\gamma_j\gamma_5\partial_i\phi
+\bar{\psi}_j\gamma_i\gamma_5\partial_j\phi\right)\nonumber\\
&&+\frac{g_3}{2}
\left(\bar{\psi}_0\gamma_0\gamma_j\gamma_5\gamma_i\partial_j\phi
-\bar{\psi}_j\gamma_j\gamma_k\gamma_5\gamma_i\partial_k\phi\right)\nonumber\\
&&-i\partial_k\bar{\psi}_i\gamma_k
-iA\left(\partial_i\bar{\psi}_0\gamma_0
+\partial_i\bar{\psi}_k\gamma_k
+\partial_k\bar{\psi}_k\gamma_i\right)\nonumber\\
&&+\frac{i}{2}(3A^2+2A+1)\left(-\partial_k\bar{\psi}_0\gamma_0\gamma_k\gamma_i
+\partial_k\bar{\psi}_j\gamma_j\gamma_k\gamma_i\right)\nonumber\\
&&-m(3A^2+3A+1)\left(\bar{\psi}_0\gamma_0\gamma_i
-\bar{\psi}_k\gamma_k\gamma_i\right)+m\bar{\psi}_i\nonumber\\
&&+i\lambda_i^{\dag}
+\frac{i}{2}(3A^2+4A+1)\lambda_0^{\dag}\gamma_0\gamma_i
+\frac{i}{2}(3A^2+2A+1)\lambda_j^{\dag}\gamma_j\gamma_i\,,
\label{poissoni}
\end{eqnarray}
and analogous equations from
$\{\theta^0_{\psi^\dag},H\}=0=\{\theta^i_{\psi^\dag},H\}$.
  The set of linear equations (\ref{poisson0}) and (\ref{poissoni}) may then
be formulated as
\begin{equation}
\lambda^\dag\mathrm M^\dag = X^\dag\,, \label{Glsys1}
\end{equation}
and analogously
\begin{equation}
\mathrm M \lambda = X\,,\label{Glsys2}
\end{equation}
where $\mathrm M$ and $\mathrm M^\dag$ are $16\times16$ matrices.
   The expressions for M and $X$ are given in the appendix.
   In the following we only consider the equations for
$\lambda$ explicitly, the equations for $\lambda^\dag$ can be
treated analogously.
   Equation (\ref{Glsys2}) cannot be solved for all the $\lambda$'s, since
\begin{equation}
\det \mathrm M=0\,.
\end{equation}
   But this was to be expected, because until now only 32
constraints have been generated [Eqs.~(\ref{zwangsbed1}) plus the
analogous constraints $\theta^0_{\psi^\dag}$ and
$\theta^i_{\psi^\dag}$].
   To get the correct number of degrees of
freedom, in total 48 constraints are needed.
   The condition of solvability of Eq.~(\ref{Glsys2}) supplies further
(secondary) constraints.
   To determine these we diagonalize M
\begin{equation}
\mathrm S^{-1}\mathrm{MS}=\left(\begin{array}{cccc}2\cdot1_{4\times4}&0&0&0\\
0&2\cdot1_{4\times4}&0&0\\
0&0&-4(1+3A+3A^2)\cdot 1_{4\times4}&0\\
0&0&0&0\end{array}\right)\,,\label{SMS}
\end{equation}
where the nonsingular matrix S is given in the appendix.
   Thus Eq.\ (\ref{Glsys2}) transforms to
\begin{equation}
\mathrm S^{-1}\mathrm{MS}\,\lambda'=\mathrm
S^{-1}X\,,\label{Glsysdiag}
\end{equation}
where \begin{equation}
\lambda'=\mathrm S^{-1}\lambda\,.
\end{equation}
   As can be seen from S$^{-1}$MS in Eq.\ (\ref{SMS}), the so far
unknown Lagrange multiplies $\lambda'_0$, $\lambda'_1$, and
$\lambda'_2$ can be determined from Eq.\ (\ref{Glsysdiag}), but
it  cannot be solved for $\lambda'_3$.
   On the other hand, secondary constraints are generated,
which read\footnote{Recall that we count the components from 0 to 3.}
\begin{equation}
\theta^4_{\psi^\dag}=(\mathrm S^{-1}X)_3=0\,.\label{sekZw}
\end{equation}
   In total we obtain eight more constraints [four from Eq.~(\ref{sekZw})
and four from the analogous procedure for $\lambda^{\dag}$], so we are
still missing eight constraints.
   To obtain these, it is necessary that the time conservation of the
secondary constraints does not lead to a determination of
$\lambda'_3$.
   Equation (\ref{sekZw}) reads in full
\begin{equation}
\theta^4_{\psi^\dag}=(\mathrm
S^{-1}X)_3=\frac{1+A}{4(1+3A+3A^2)}\gamma_0\gamma_1\gamma_3
\left[(1+3A)\gamma_0X_0+(1+A)\gamma_iX_i\right]=0\,.
\end{equation}
This condition is equivalent to
\begin{equation}
{\theta^4_{\psi^\dag}}'=(1+3A)\gamma_0X_0+(1+A)\gamma_iX_i=0\,.\label{sekZw2}
\end{equation}
Substituting Eqs.\ (\ref{X0})--(\ref{Xi}) for $X$, Eq.\
(\ref{sekZw2}) can be written as
\begin{equation}
{\theta^4_{\psi^\dag}}'=(\mathcal F-\pi)\xi_1
+\xi_2+\xi_3=0,
\label{sekZwausgeschr}
\end{equation}
where
\begin{eqnarray*}
\xi_1&=&i\gamma_5\{
\left[-(1+3A)(g_1+2g_2-g_3)+3(1+A)(g_2-g_3)\right]\gamma_0\psi_0\nonumber\\
&&+\left[(1+3A)(g_2-g_3)+(1+A)g_1+3(1+A)g_3\right]
\gamma_i\psi_i\}\,,\\
\xi_2&=& i\gamma_5\{
\left[-(1+3A)(g_1+g_3)-(1+A)g_2+3(1+A)g_3\right]\gamma_i\psi_0\nonumber\\
&&+\left[2g_2+2(1+A)g_1\right]\gamma_0\psi_i
+\left[2g_3-(1+A)g_2+(1+A)g_1\right]\gamma_0\gamma_i\gamma_j\psi_j
\}\partial_i\phi\,,\\
\xi_3&=&4(2A+1)\gamma_0\partial_i\psi_i
-4(2A^2+3A+1)\gamma_i\partial_i\psi_0
-4A(2A+1)\gamma_0\gamma_i\gamma_j\partial_i\psi_j\nonumber\\
&&+2im\left[3(1+A)(1+2A)\psi_0-(6A^2+5A+1) \gamma_0\gamma_i\psi_i\right]\,.
\end{eqnarray*}
   To conserve this constraint in time, the Poisson bracket of
${\theta^4_{\psi^\dag}}'$ with the Hamiltonian must be equal to zero:
\begin{equation}
\{{\theta^4_{\psi^\dag}}',H\}\stackrel{!}{=}0\,.
\label{pbtheta4ph}
\end{equation}
   As explained above, to obtain further constraints,
we cannot allow for $\lambda'_3$ to be solvable from
Eq.\ (\ref{pbtheta4ph}).
   To satisfy this condition we require that
$\psi_3'=\left( S^{-1}\psi\right)_3$ must not appear
in Eq.\ (\ref{sekZwausgeschr}).
   The following conclusions can be drawn from this condition:
\begin{enumerate}
\item Since $\mathcal F$ contains the combination $\psi_3'$, $\xi_1$ of
Eq.\ (\ref{sekZwausgeschr}) has to disappear.
   This results in
\begin{eqnarray}
-(1+3A)(g_1+2g_2-g_3)+3(1+A)(g_2-g_3)&=&0\,,\\
(1+3A)(g_2-g_3)+(1+A)g_1+3(1+A)g_3&=&0\,,
\end{eqnarray}
which is fulfilled, if and only if
\begin{eqnarray}
g_2&=&Ag_1\,,\label{g2ausquant}\\
g_3&=&-\frac{1}{2}(1+2A+3A^2)g_1\,.\label{g3ausquant}
\end{eqnarray}
\item
   Secondly, all terms containing $\psi_3'$ in $\xi_2$ have to
disappear.
   But this is automatically fulfilled, if Eqs.\ (\ref{g2ausquant}) and
(\ref{g3ausquant}) hold.
\item
The remaining terms do not contain the combination $\psi'_3$, so
no more conditions occur.
\end{enumerate}
   Thus, if Eqs.\ (\ref{g2ausquant}) and (\ref{g3ausquant}) hold, we
get eight more constraints, namely
\begin{equation}
\theta^5_{\psi^\dag}=\{\theta^4_{\psi^\dag},H\}=0,
\end{equation}
and analogous expression for $\theta^5_{\psi}$.
   Demanding the time independence of $\theta^5_{\psi^\dag}$ (and
$\theta^5_{\psi}$) no more constraints are generated and all remaining
$\lambda'_3$ multipliers are determined.
   Thus the chain of constraints terminates at this point and the correct
number of physical degrees of freedom has been generated.

   It is interesting to note that, after inserting Eqs.\
(\ref{g2ausquant}) and (\ref{g3ausquant}), the total Lagrangian of
Eq.~(\ref{Lges}) fulfills the point invariance of
Eqs.~(\ref{punkt1}) and (\ref{punkt2}).
   Thus, a suitable field redefinition in the form of Eq.\ (\ref{punkt1})
transforms the Lagrangian from a general $A$ to a fixed $A$, say, $A=-1$.

\subsection{Constraint analysis in an arbitrary dimension $n$}
   The calculation for $n$ dimensions and $A=-1$ is analogous to the
one of the previous subsection.
   In fact, it is simpler, since the resulting sets of linear equations
corresponding to Eqs.~(\ref{Glsys1}) and (\ref{Glsys2}) already
have diagonal form.
   For $A=-1$ the Lagrangian does not explicitly
depend on the space-time dimension $n$.
   Thus we obtain the same
result as we get when substituting $A=-1$ into Eqs.\
(\ref{g2ausquant}) and (\ref{g3ausquant}), namely,
\begin{equation}
g_2=g_3=-g_1\,.
\end{equation}
   With these conditions the chain of constraints terminates at the
correct number of degrees of freedom.
   To obtain the equivalent conditions for a general $A$ and an arbitrary
dimension $n$, we use the following transformation
\begin{equation}
\psi^{\mu}\rightarrow
\left(g^{\mu\nu}-\frac{A+1}{n-2}\gamma^{\mu}\gamma^{\nu}\right)\psi_{\nu}\,.
\end{equation}
   This leads to the following relations among the coupling constants
\begin{eqnarray}
g_2(A)&=&A g_1\,,\\
g_3(A)&=&-\frac{1+2A+A^2(n-1)}{n-2}\label{g2g3ndim}\,.
\end{eqnarray}
   Of course, for $n=4$ we reproduce the results of the previous
subsection.

   The requirement of the consistency of the theory automatically
results in the invariance under the point transformation.
   On the other hand, demanding the invariance under the point transformation
alone is not sufficient to obtain the relations of  Eqs.\
(\ref{g2ausquant}) and (\ref{g3ausquant}) as will be shown in the
next section.

\section{Conditions for point invariance}
   To determine the general conditions for the coupling constants,
that follow from requiring the point invariance of the Lagrangian,
we apply the point transformation in four dimensions to the
interaction term of the Lagrangian:
\begin{eqnarray}
\psi^{\mu}&\rightarrow&\psi^{\mu}
+a\gamma_{\mu}\gamma_{\beta}\psi^{\beta}\,,\nonumber\\
A&\rightarrow&A'=\frac{A-2a}{1+4a}\,,\qquad
a\neq-\frac{1}{4}\,.\label{punkttrafo}
\end{eqnarray}
The change of the Lagrangian under this transformation,
$\Delta\mathcal L=\mathcal L'-\mathcal L$, is then set to
zero.\footnote{In principle $\Delta\mathcal L$ could also be a
total derivative, but this does not apply in the present case.}
   We obtain for $\Delta\mathcal L$
\begin{equation}
\Delta\mathcal L=\Delta\mathcal L_1+\Delta\mathcal
L_2+\Delta\mathcal L_3\,,
\end{equation}
where
\begin{eqnarray}
\Delta\mathcal L_1&=&-g_1
a\bar{\psi}^{\mu}[(u_{\mu}\gamma_{\nu}+u_{\nu}\gamma_{\mu})\gamma_5
+(1+a)\gamma_{\mu}\slashed
u \gamma_5\gamma_{\nu}]\psi^{\nu}\,, \nonumber\\
\Delta\mathcal
L_2&=&-\frac{1}{2}\bar{\psi}^{\mu}\{[g_2(A')(1+4a)-g_2(A)]
(\gamma_{\mu}u_{\nu}+u_{\mu}\gamma_{\nu})\gamma_5
-2a(1+4a)g_2(A')\gamma_{\mu}\slashed
u \gamma_5\gamma_{\nu}\}\psi^{\nu}\,,\nonumber\\
\Delta\mathcal
L_3&=&-\frac{1}{2}[g_3(A')(1+4a)^2-g_3(A)]
\bar{\psi}^{\mu}\gamma_{\mu}\slashed
u \gamma_{5}\gamma_{\nu}\psi^{\nu}\,.\nonumber
\end{eqnarray}
   To fulfill $\Delta\mathcal L=0$ we obtain the following conditions
\begin{eqnarray}
g_1 a+\frac{1}{2}[g_2(A')(1+4a)-g_2(A)]&=&0\label{I}\,,\\
g_1a(1+a)-a(1+4a)g_2(A')+\frac{1}{2}[g_3(A')(1+4a)^2-g_3(A)]&=&0\,.\label{II}
\end{eqnarray}
   These functional equations for $g_2(A)$ and $g_3(A)$ can be
solved.
   The solutions are
\begin{eqnarray}
g_2(A)&=&g_1\left[2z_2+(1+4z_2)A\right]\,,\label{g2}\\
g_3(A)&=&2g_1\left[z_3+\left(\frac{1}{2}-z_2+4z_3\right)A
+\left(\frac{1}{4}+4z_3-2z_2\right)A^2\right]\,,\label{g3}
\end{eqnarray}
where $z_2$ and $z_3$ are arbitrary (see also Ref.~\cite{Tang:1996sq}).
   We compare these with the relations that follow from consistency,
Eqs.\ (\ref{g2ausquant}) and (\ref{g3ausquant}). As we can see,
the latter are more stringent, because the parameters that are
arbitrary in Eqs.\ (\ref{g2}) and (\ref{g3}), are fixed to
\begin{eqnarray}
z_2&=&0\,,\\
z_3&=&-\frac{1}{4}\,,
\end{eqnarray}
when requiring consistency of the theory.

\section{Summary}

   We have considered the Hamilton formalism for a system of spin-3/2
particles interacting with pseudoscalars.
   The Lorentz-invariant formulation of a field theory for spin-3/2
particles necessarily leads to the introduction of unphysical
degrees of freedom.
   In order to obtain the right number of degrees of
freedom some constraints have to be imposed.

   In the present work we have considered the Rarita-Schwinger formulation
for spin-3/2 particles and have analyzed the constraint structure
for the lowest-order $\pi\Delta$ interaction terms.
   The requirement of the consistency of the corresponding effective
field theory in the sense of having the right number of degrees of
freedom has led to non-trivial constraints among the three
coupling constants of the lowest-order $\pi\Delta$ interaction.
   As a result of these constraints the total Lagrangian is
invariant under the so-called point transformation, guaranteeing
that the physical quantities are independent of the off-shell
parameter $A$.
   On the other hand, demanding the invariance under
the point transformation alone is less stringent and produces only
a class of relations among the coupling constants.
   We conclude that the analysis of the constraint structure is an
important ingredient in the construction of the
most general effective field theory including particles with spin
$S\geq 1$.
   In particular, as a rule it leads to a reduction in the number
of free parameters of the Lagrangian.

\acknowledgments

The authors would like to thank D.~Djukanovic for useful comments
on the manuscript.
   The work of J.~G.~was supported by the Deutsche Forschungsgemeinschaft
(contract SCHE 459/2-1).

\appendix
\section{M, $X$ and S}
 The left-hand
side of Eq.\ (\ref{Glsys2}) is given by
\begin{eqnarray}
(\mathrm M
\lambda)_0&=&-\underbrace{(3A^2+6A+3)}_{:=B}\lambda_0
+\underbrace{(3A^2+4A+1)}_{:=C}\gamma_0\gamma_i\lambda_i\,,\\
(\mathrm
M\lambda)_i&=&2\lambda_i+(3A^2+4A+1)\gamma_0\gamma_i\lambda_0
 +\underbrace{(3A^2+2A+1)}_{:=D}\gamma_i\gamma_j\lambda_j\,.
\end{eqnarray}
Thus the matrix M reads
\begin{equation}
\mathrm M=\left(\begin{array}{cccc}-B&C\gamma_0\gamma_1&
C\gamma_0\gamma_2&C\gamma_0\gamma_3\\
C\gamma_0\gamma_1&2-D&D\gamma_1\gamma_2&D\gamma_1\gamma_3\\
C\gamma_0\gamma_2&D\gamma_2\gamma_1&2-D&D\gamma_2\gamma_3\\
C\gamma_0\gamma_3&D\gamma_3\gamma_1&D\gamma_3\gamma_2&2-D
\end{array}\right)\,.
\end{equation}
The components of the vector $X$ on the right-hand side of Eq.\
(\ref{Glsys2}) are given by
\begin{eqnarray}
X_0&=&i(\mathcal F-\pi)\gamma_5\left[(g_1+2g_2-g_3)\psi_0
-(g_2-g_3)\gamma_0\gamma_i\psi_i\right]\nonumber\\
&&+i\partial_i\phi\gamma_5\left[(g_1+g_3)\gamma_0\gamma_i\psi_0
+g_2\psi_i+g_3\gamma_i\gamma_j\psi_j\right]\nonumber\\
&&-2A\partial_i\psi_i+(3A^2+2A-1)\gamma_0\gamma_i\partial_i\psi_0
+(3A^2+2A+1)\gamma_i\gamma_j\partial_i\psi_j\nonumber\\
&&-6imA(A+1)\gamma_0\psi_0+2im(3A^2+3A+1)\gamma_i\psi_i\label{X0}\,,\\
X_i&=&-i(\mathcal F-\pi)\left[g_1\gamma_5\psi_i-(g_2-g_3)
\gamma_0\gamma_5\gamma_i\psi_0
+g_3\gamma_i\gamma_5\gamma_j\psi_j\right]\nonumber\\
&&+ig_1\partial_j\phi\gamma_0\gamma_5\gamma_j\psi_i
+ig_2\left(\partial_i\phi\gamma_5\psi_0
+\partial_i\phi\gamma_0\gamma_5\gamma_j\psi_j
+\partial_j\phi\gamma_0\gamma_5\gamma_i\psi_j\right)\nonumber\\
&&+2\gamma_0\gamma_k\partial_k\psi_i
+2A\left(-\partial_i\psi_0+\gamma_0\gamma_k\partial_i\psi_k
+\gamma_0\gamma_i\partial_k\psi_k\right)\nonumber\\
&&-(3A^2+2A+1)\left(-\gamma_i\gamma_k\partial_k\psi_0
+\gamma_0\gamma_i\gamma_k\gamma_j\partial_k\psi_j\right)\nonumber\\
&&-2im(3A^2+3A+1)\left(\gamma_i\psi_0+\gamma_0\gamma_i\gamma_k\psi_k\right)
-2im\gamma_0\psi_i\label{Xi}\,.
\end{eqnarray}
The matrix S of Eq.\ (\ref{SMS}) is given by
\begin{equation}
\mathrm{S}=\left(\begin{array}{cccc}0&0&
\frac{3(1+A)}{1+3A}\gamma_3\gamma_1&\frac{1+3A}{1+A}\gamma_3\gamma_1\\
\gamma_3\gamma_0&\gamma_2\gamma_0&\gamma_0\gamma_3&\gamma_0\gamma_3\\
0&\gamma_1\gamma_0&\gamma_0\gamma_1\gamma_2\gamma_3&
\gamma_0\gamma_1\gamma_2\gamma_3\\
\gamma_1\gamma_0&0&\gamma_1\gamma_0&\gamma_1\gamma_0\end{array}\right)\,.
\end{equation}

\end{document}